\documentclass[letterpaper, 10 pt, conference]{ieeeconf}  

\IEEEoverridecommandlockouts                              

\usepackage{amsthm}
\usepackage{graphics} 
\usepackage{epsfig} 
\usepackage{times} 
\usepackage{amsmath} 
\usepackage{amssymb}  

\usepackage{biblatex}
\usepackage{bm}
\usepackage{xcolor}
\usepackage{graphicx}
\usepackage{algorithm}
\usepackage[noend]{algpseudocode}
\usepackage{booktabs}
\usepackage{siunitx}
\usepackage{hyperref}

\usepackage{caption}
\usepackage{lipsum}
\usepackage{multirow}
\usepackage{diagbox} 
\usepackage{subcaption}

\newcommand{\R}{\mathbb{R}}

\newcommand{\sigi}{\sigma_i}
\newcommand{\sigj}{\sigma_j}

\newtheorem{definition}{\bf Definition}
\newtheorem{assumption}{\bf Assumption}
\newtheorem{theorem}{\bf Theorem}
\newtheorem{lemma}{\bf Lemma}
\newtheorem{corollary}{\bf Corollary}
\newtheorem{remark}{\bf Remark}

\title{\LARGE \bf

Distributionally Robust Cascading Risk Quantification in Multi-Agent Rendezvous: Effects of Time Delay and Network Connectivity
}

\author{Vivek Pandey and Nader Motee 
\thanks{
 V. Pandey and N. Motee are with the Department of Mechanical Engineering and Mechanics, Lehigh University, Bethlehem, PA, 18015, USA. {\tt\small \{vkp219,  motee\}@lehigh.edu}.\endgraf
}
}

\begin{document}

\maketitle

\thispagestyle{plain}
\pagestyle{plain}

\begin{abstract} 
Achieving safety in autonomous multi-agent systems, particularly in time-critical tasks like rendezvous, is a critical challenge. In this paper, we propose a distributionally robust risk framework for analyzing cascading failures in multi-agent rendezvous. To capture the complex interactions between network connectivity, system dynamics, and communication delays, we use a time-delayed network model as a benchmark. We introduce a conditional distributionally robust functional to quantify cascading effects between agents, utilizing a bi-variate normal distribution. Our approach yields closed-form risk expressions that reveal the impact of time delay, noise statistics, communication topology, and failure modes on rendezvous risk. The insights derived inform the design of resilient networks that mitigate the risk of cascading failures. We validate our theoretical results through extensive simulations, demonstrating the effectiveness of our framework.
\end{abstract}


\section{Introduction}
Networked systems are ubiquitous and play a crucial role in various engineering applications, from power networks \cite{dorfler2012synchronization} to autonomous vehicle platoons \cite{liu2021risk}. However, their performance is often hindered by communication challenges and external disturbances, which can push the system away from its optimal state, leading to inefficiencies or even complete network failure. Such disruptions are frequently observed in power networks \cite{Somarakis2020power}, supply chains \cite{bertsimas1998air}, and financial systems \cite{acemoglu2015systemic}, where cascading failures can result in widespread breakdowns.

The complexity of networked systems also introduces fragility, making them susceptible to systemic failures. In consensus networks \cite{olfati2004consensus}, where multiple agents coordinate to reach a common decision, factors such as time delays and environmental noise can cause deviations from consensus. These uncertainties have drawn significant attention to the risk analysis of complex dynamical systems \cite{somarakis2018risk,Somarakis2016g,Somarakis2019g}. Understanding and quantifying these risks are essential for robust system design, as the failure of a single agent can propagate through the network, triggering a cascading effect that compromises overall system stability \cite{zhang2018cascading,zhang2019robustness}.

In the realm of multi-agent consensus networks, several risk assessment methods, including Value at Risk (VaR) \cite{rockafellar2000optimization} and Conditional Value at Risk (CVaR) \cite{rockafellar2002conditional}, have been utilized to quantify the risk of single failures \cite{Somarakis2016g, Somarakis2017a, Somarakis2019g} and cascading failures \cite{liu2021risk,liu2023cr_risk_first_order}. These methods typically assume known probability distributions of uncertainties, which may not be readily available in practical, real-world settings. The existence of distributional ambiguity—where the true probability distribution is either unknown or fluctuates over time—calls for alternative risk evaluation techniques that do not rely on precise distributional assumptions.

To address this challenge, we propose a distributionally robust risk framework that explicitly accounts for ambiguity in probability distributions. Instead of assuming a single known distribution, we define an ambiguity set—a collection of plausible distributions consistent with prior knowledge or observed data. This allows us to evaluate risk measures that remain valid under varying distributions, thereby improving the reliability of risk-aware decision-making in networked systems.

{\it Our Contributions:}
Building on the distributionally robust risk framework developed in our previous work on platoons of vehicles \cite{pandey2023dr_risk_second_order}, this paper takes the next step by applying the framework to the multi-agent rendezvous problem. We begin by using a conditional distributionally robust functional to define risk, which is then quantified through the steady-state statistics of a time-delayed multi-agent rendezvous system. Specifically, we derive an explicit formula for the distributionally robust cascading risk. Through extensive simulations, we present the cascading risk profiles in a rendezvous scenario, exploring how these profiles evolve under various conditions. Through theoretical analysis and simulations, we demonstrate that higher network connectivity does not necessarily lead to reduced risk.

\section{Preliminaries and Notation}
We define the non-negative orthant of the Euclidean space \(\mathbb{R}^n\) as \(\mathbb{R}_{+}^n\). The standard Euclidean basis in \(\mathbb{R}^n\) is given by \(\{\bm{e}_1, \dots, \bm{e}_n\}\), and the all-ones vector is denoted by \(\bm{1}_n = [1, \dots, 1]^T\). We denote by \(\text{S}_+^n,\) the cone of \(n \times n\) positive semidefinite matrices. For matrices \(A, B \in \mathbb{R}^{n \times n}\), we write \(A \succeq B\) to indicate that \(A - B\) is positive semidefinite.

\vspace{0.1cm}
{\it Spectral Graph Theory:} A weighted graph is characterized by \(\mathcal{G} = (\mathcal{V}, \mathcal{E}, \omega)\), where \(\mathcal{V}\) is the set of nodes, \(\mathcal{E}\) is the set of edges, and \(\omega: \mathcal{V} \times \mathcal{V} \to \mathbb{R}_{+}\) is the weight function that assigns a non-negative value to each edge. Two nodes are said to be directly connected if and only if \((i, j) \in \mathcal{E}\).

\begin{assumption}  \label{asp:connected}
    The graph under consideration is assumed to be connected. Furthermore, for all \(i, j \in \mathcal{V}\), the following conditions hold:
    \begin{itemize}
        \item \(\omega(i,j) > 0\) if and only if \((i,j) \in \mathcal{E}\).
        \item \(\omega(i,j) = \omega(j,i)\), i.e., that the edges are undirected.
        \item \(\omega(i,i) = 0\), implying that the edges are simple.
    \end{itemize}
\end{assumption}

The Laplacian matrix of the graph \(\mathcal{G}\) is an \(n \times n\) matrix \(L = [l_{ij}]\) with elements given by

\[
    l_{ij} := \begin{cases}
        -k_{i,j}  & \text{if } i \neq j, \\
        k_{i,1} + \dots + k_{i,n}  & \text{if } i = j,
    \end{cases}
\]
where \(k_{i,j} := \omega(i,j)\). The Laplacian matrix is symmetric and positive semi-definite \cite{van2010graph}. Assumption \ref{asp:connected} implies that the smallest eigenvalue of the Laplacian is zero, with an algebraic multiplicity of one. The eigenvalues of \(L\) can be ordered as~\(
    0 = \lambda_1 < \lambda_2 \leq \dots \leq \lambda_n.
\)
The eigenvector corresponding to the eigenvalue \(\lambda_k\) is denoted by \(\bm{q}_k\). By letting \(Q = [\bm{q}_1 | \dots | \bm{q}_n]\), we can express the Laplacian matrix as \(L = Q \Lambda Q^T\), where \(\Lambda = \text{diag}[0, \lambda_2, \dots, \lambda_n]\). We normalize the Laplacian eigenvectors so that \(Q\) becomes an orthogonal matrix, i.e., \(Q^T Q = Q Q^T = I_n\), with \(\bm{q}_1 = \frac{1}{\sqrt{n}} \bm{1}_n\).

\vspace{0.1cm}

{\it Probability Theory:} Let \(\mathcal{L}^{2}(\mathbb{R}^{q})\) denote the space of \(\mathbb{R}^q\)-valued random vectors \(\bm{z} = [z^{(1)}, \dots ,z^{(q)}]^T\) defined on a probability space \((\Omega, \mathcal{F}, \mathbb{P})\) with finite second moments. A normal random variable \(\bm{y} \in \mathbb{R}^{q}\) with mean \(\bm{\mu} \in \mathbb{R}^{q}\) and covariance matrix \(\Sigma \in \mathbb{R}^{q \times q}\) is expressed as \(\bm{y} \sim \mathcal{N}(\bm{\mu}, \Sigma)\). The error function, denoted \(\text{erf} : \mathbb{R} \to (-1, 1)\), is given by

\[
\text{erf}(x) = \frac{2}{\sqrt{\pi}} \int_{0}^{x} e^{-t^2} \, \text{d} t.
\]

Additionally, we use the standard notation \(\text{d} \bm{\xi}_t\) in the context of formulating stochastic differential equations.

\section{Problem Statement}\label{problemstatement}

We focus on time-delayed linear consensus networks, which have broad applications in engineering, including clock synchronization in sensor networks, rendezvous in space or time, and heading alignment in swarm robotics. For more details, we refer to \cite{ren2007information, olfati2007consensus}. As a motivating example, we explore the time-delayed rendezvous problem, where the goal is for a group of agents to meet simultaneously at a pre-specified location known to all. 

In this scenario, agents lack prior knowledge of the meeting time, which may need to be adjusted in response to unforeseen emergencies or external uncertainties \cite{ren2007information}. Thus, the agents must reach a consensus on the rendezvous time. This consensus is achieved by each agent \( i = 1, \dots, n \) creating a state variable, say \( x_i \in \mathbb{R} \), which represents its belief about the rendezvous time. Initially, each agent sets its belief to the time it prefers for the rendezvous. The dynamics of each agent's belief evolve over time according to the following stochastic differential equation:

\begin{equation} \label{eqn:individual_dynamics}
\text{d} x_i(t) = u_i(t) \, \text{d} t + b \, \text{d} w_i(t),
\end{equation}

for all \( i = 1, \dots, n \).

Each agent's control input is denoted by \( u_i \in \mathbb{R} \). Uncertainty in the network propagates as additive stochastic noise, with its magnitude uniformly scaled by the diffusion coefficient \( b \in \mathbb{R} \). The influence of environmental uncertainties on agent dynamics is modeled using independent Brownian motions \( w_1, \dots, w_n \). 

In practical scenarios, agents often experience time delays in accessing, computing, or sharing their own and neighboring agents' state information \cite{ren2007information}. To account for this, we assume that all agents experience a uniform time delay \( \tau \in \mathbb{R}_{+} \). 

The control inputs are determined through a negotiation process, where agents form a linear consensus network over a communication graph. The control law governing the system is given by

\begin{equation} \label{eq:feedback}
    u_i(t) = \sum_{j = 1}^{n} k_{ij} \left(x_j(t-\tau) - x_i(t-\tau) \right),
\end{equation}

where \( k_{ij} \in \mathbb{R}_{+} \) are nonnegative feedback gains.

Let us denote the state vector as \( \bm{x}_t = \left[x_1(t), \dots, x_n(t) \right]^T, \) and the vector of exogenous disturbances as \( \bm{w}_t = \left[w_1(t), \dots, w_n(t) \right]^T \). The dynamics of the resulting closed-loop network can be expressed as a linear consensus network governed by following the stochastic differential equation:

\begin{equation} \label{eqn: network_dynamics}
    \text{d}  \bm{x}_t = -L \, \bm{x}_{t-\tau}\, \text{d}  t + B \, \text{d}  \bm{w}_t,
\end{equation}

for all \( t \geq 0 \), where the initial function \( \bm{x}_t = \phi(t) \) is deterministically given for \( t \in [-\tau, 0] \), and \( B = b I_n \). 

The underlying coupling structure of the consensus network \eqref{eqn: network_dynamics} is represented by a graph \( \mathcal{G} \) that satisfies Assumption \ref{asp:connected}, with the associated Laplacian matrix \( L \). We assume that the communication graph \( \mathcal{G} \) remains time-invariant, ensuring that the network of agents reaches consensus on a rendezvous time before executing motion planning to reach the designated meeting location. Once consensus is achieved, a properly designed internal feedback control mechanism guides each agent toward the rendezvous location.


\begin{assumption} \label{asp:stable}
\cite{Somarakis2019g}The time-delay is assumed to satisfy
\[
\tau < \frac{\pi}{2 \lambda_n},
\]
which guarantees the stability of the network \eqref{eqn: network_dynamics}.
\end{assumption}

When there is no noise, i.e., \(b = 0\), it is known from \cite{olfati2004consensus} that, under Assumptions \ref{asp:connected} and \ref{asp:stable}, the state of every agent converges to the average of the initial states, namely,
\[
\frac{1}{n}\bm{1}_n^T \bm{x}_0.
\]
In contrast, in the presence of input noise, the state variables fluctuate around the instantaneous average \(\frac{1}{n}\bm{1}_n^T \bm{x}_t\). 

To quantify the quality of the rendezvous and capture its fragility features, we define the vector of observables
\begin{equation} \label{eq:observables}
    \bm{y}_t = M_n\, \bm{x}_t,
\end{equation}
where the centering matrix \(M_n = I_n - \frac{1}{n}\bm{1}_n \bm{1}_n^T\) ensures that \(\bm{y}_t = \bigl(y_1(t), \dots, y_n(t)\bigr)^T\) measures the deviation of each agent from the consensus value.

Under Assumption \ref{asp:connected}, one mode of the network described by \eqref{eqn: network_dynamics} is marginally stable, corresponding to the zero eigenvalue of the Laplacian matrix \(L\). This marginally stable mode is unobservable from the output \eqref{eq:observables}, ensuring that \(\bm{y}_t\) remains bounded in the steady state. In the absence of noise, we have \(\bm{y}_t \rightarrow \bm{0}\) as \(t \rightarrow \infty\). However, when noise is present, it excites the observable modes, leading the output to fluctuate around zero. Consequently, the agents cannot agree on an exact rendezvous time, and a practical solution is to permit a tolerance interval within which consensus is deemed acceptable.
\begin{definition}[c-Consensus]
For a given tolerance level \( c \in \mathbb{R}_+ \), the consensus network \eqref{eqn: network_dynamics} achieves \emph{c-consensus} if it can tolerate certain degrees of disagreement such that
\[
\lim_{t \to \infty} \mathbb{E}[|\bm{y}_t|] \leq c \mathbf{1}_n.
\]
\end{definition}
 This implies that all agents agree on all points in the set \( \{ x \in \mathbb{R}^n \mid  \lvert M_n x \rvert \leq c \mathbf{1}_n \}. \) However, certain events can cause the average deviation of agent \( i \)'s observable from zero to exceed the threshold \( c \), a condition defined as a systemic failure.
\begin{definition}[Systemic Failure]
For a given tolerance level \( c \in \mathbb{R}_+ \), an agent with observable \( y^{(i)}_t \) is prone to \emph{systemic failure} if 
\[
\lim_{t \to \infty} \mathbb{E}[|y^{(i)}_t|] > c.
\]
\end{definition}

The primary challenge addressed in this work is to quantify the distributionally robust cascading risk in multi-agent rendezvous. Specifically, we aim to determine how the interplay between the graph Laplacian, time-delay, and noise statistics influences the risk of failing to reach \(c\)-consensus, particularly when some agents have already failed. To this end, we develop a systemic risk framework that evaluates cascading large fluctuations based on the steady-state statistics of the closed-loop system.

\section{Preliminary Results}\label{prelims}

In this section, we analyze the steady-state statistics of network observables under external disturbances and time-delay. We introduce systemic events, which quantify large fluctuations of observables from the mean, and develop a distributionally robust risk framework to assess the risk of cascading failures under distributional ambiguity.

\subsection{Steady-State Statistics of Observables}

It is known from \cite{Somarakis2019g} that if Assumptions \ref{asp:connected} and \ref{asp:stable} hold, then as \(t \rightarrow \infty\), the steady-state observables defined in \eqref{eq:observables} converge in distribution to a normal distribution:
\[
\bar{\bm{y}} \sim \mathcal{N}(0, \Sigma),
\]
where \(\Sigma \in \text{S}^{n}_+\) is the covariance matrix.



\begin{lemma} \label{lem:sigma_y_steady}
    The steady-state covariance matrix \(\Sigma\) of the observables \eqref{eq:observables} is given by
    \begin{equation} \label{eq:sigma_y}
        \Sigma = \frac{1}{2} b^2 M_n Q \bar{\Lambda} Q^T M_n,
    \end{equation}
    where \(\bar{\Lambda} = \operatorname{diag}(0, \frac{\cos (\lambda_2 \tau)}{\lambda_2 (1 - \sin (\lambda_2 \tau))}, \dots, \frac{\cos (\lambda_n \tau)}{\lambda_n (1 - \sin (\lambda_n \tau))})\) is a diagonal matrix.
\end{lemma}

For simplicity, we denote the $(i,j)$-th element of $\Sigma$ as $\sigma_{ij}$, and the diagonal elements as $\sigma_{ii} = \sigma_i^2$. Lemma \ref{lem:sigma_y_steady} illustrates the influence of time delays and network topology on the steady-state statistics of observables. This relationship is pivotal in demonstrating the network's role in risk quantification, with the help of the following result. 

\begin{figure}[t]
    \centering
    \includegraphics[width=0.96\columnwidth]{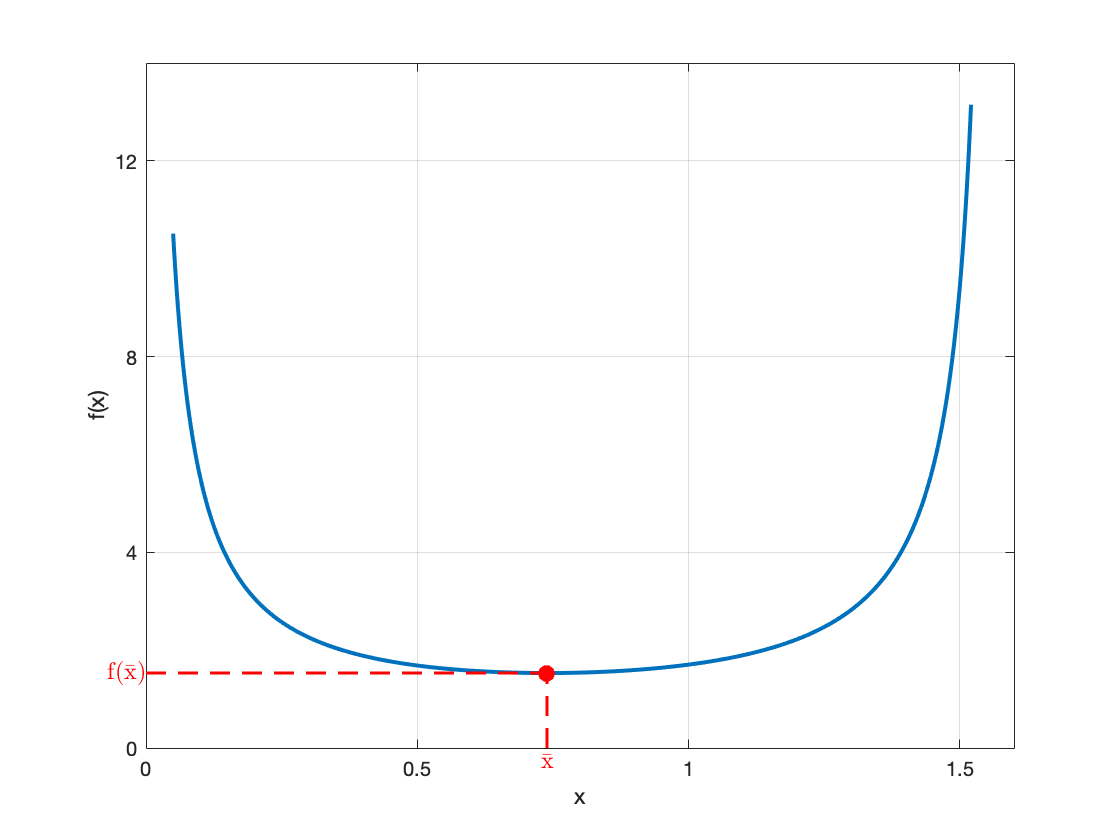} 
    \caption{Graph of \(f(x) = \frac{\cos(x)}{2x (1- \sin(x))}\).}
    \label{fig:fx_graph}
\end{figure}
\begin{lemma}\label{lem:Sigma_L_monotonicity}
    Consider two networks with graph Laplacians \(L_1\) and \(L_2\), differing by uniform edge weights such that \(L_1 \preceq L_2\). Assuming a constant time delay \(\tau\) that satisfies Assumption \ref{asp:stable}, the covariance matrices \(\Sigma_1\) and \(\Sigma_2\) satisfy:
    \[
    \Sigma_1 \succeq \Sigma_2 \quad \text{if} \quad \lambda_{n,2} \leq \bar{\lambda},
    \]
    \vspace{-0.5cm}
    \[
    \Sigma_1 \preceq \Sigma_2 \quad \text{if} \quad \lambda_{n,1} \geq \bar{\lambda},
    \]
    where \(\lambda_{n,k}\) denotes the spectral radius of \(L_k\) for \(k = 1, 2\), and \(\bar{\lambda}\) is the value of \(\lambda\) that minimizes the function \(f(\lambda \tau) \) as depicted in Fig. \ref{fig:fx_graph}.
\end{lemma}
\begin{remark}
Lemma \ref{lem:Sigma_L_monotonicity} implies that graphs \( \mathcal{G}_1 \) and \( \mathcal{G}_2 \) have the same topology, differing only by a uniform scaling factor \( \kappa \) on their edge weights, where \( 0 < \kappa \leq 1 \). This condition is trivially satisfied by path graphs.
\end{remark}

Lemma \ref{lem:Sigma_L_monotonicity} illustrates the non-monotonic dependence of the covariance matrix \(\Sigma\) on both the eigenvalues of the Laplacian and the time-delay. This intricate interplay between time-delay and eigenvalues plays a crucial role in shaping the risk, which is further explored in the subsequent sections.

To quantify the cascading effect of one random variable on another, it is essential to understand their joint probability distributions. We now present the relevant joint distribution.

\begin{lemma}\label{lem:bivariate_normal}
    The joint probability distribution function of the normal random variables \(y_i\) and \(y_j\) is given by \(p(y_j, y_i)\), where
    \begin{equation}\label{eqn:bivariate_normal_distribution}
        p(y_j, y_i) = \frac{1}{2 \pi \rho' \sigi \sigj} \exp\left(- \frac{y_j^2}{2\sigj^2} - \frac{\left(y_i - \rho \frac{\sigi y_j}{\sigj}\right)^2}{2 \rho'^2 \sigi^2} \right)
    \end{equation}
    and \(\rho = \rho_{ij}\) denotes the correlation between the random variables \(y_i\) and \(y_j\), while \(\rho' = \sqrt{1 - \rho^2}\).
\end{lemma}

\subsection{Systemic Events}
A \textit{systemic event} is a large deviation of observables that may trigger cascading failures.
The set of such undesirable outcomes is called the systemic set, denoted as \( U \subset \mathbb{R} \). In a probability space \( (\Omega, \mathcal{F}, \mathbb{P}) \), for a random variable \( y: \Omega \rightarrow \mathbb{R} \), the systemic events are defined as:
\(
\{ \omega \in \Omega \mid y(\omega) \in U. \}
\)
To monitor how close \( y \) is to entering the systemic set \( U \), we define a family of nested super-sets \( \{ U_{\delta} \mid \delta \in [0, \infty) \} \) of \( U \) that satisfy:
\begin{itemize}
    \item \textit{Nested Property}: \( U_{\delta_2} \subseteq U_{\delta_1} \) whenever \( \delta_1 < \delta_2 \),
    \item \textit{Limit Condition}: \( \lim_{\delta \to \infty} U_{\delta} = U \).
\end{itemize}

These super-sets \( U_{\delta} \) can be adjusted to cover a neighborhood around \( U \), creating "alarm zones." As we will see in the next sections, \(\delta\) quantifies magnitude of deviation of agent's observable from mean.

\subsection{Distributionally Robust Risk Framework}  \label{risk}

To quantify distributionally robust risk, it is essential to acknowledge that the true probability distribution may be unknown or uncertain. To address this, we propose a framework that considers a family of probability distributions, rather than relying on a single known distribution. This approach captures the worst-case scenario across all possible distributions, providing a more resilient measure of risk under uncertainty.

To develop a risk measure that is robust to uncertainty in the probability distribution, we define a set of probability distributions using the notion of an ambiguity set, as given in Definition~\ref{def: ambig_set}.

\begin{definition}\label{def: ambig_set}
\textbf{Ambiguity Set of Probability Measures} \cite{shapiro2022}:  
Let \( (\Omega, \mathcal{F}) \) be a measurable space. The ambiguity set \( \mathcal{M} \) is defined as a nonempty family of probability measures on \( (\Omega, \mathcal{F}) \) that lie within a specified ball centered at a reference probability measure, with the radius determined by a chosen metric.  
\end{definition}

Building upon the ambiguity set, we now define a conditional distributionally robust functional, which forms the foundation for our distributionally robust risk measure. This functional captures the worst-case expectation of a random variable, conditioned on available information, across all probability measures within the ambiguity set.
 
\begin{definition}\label{def: condn_drf}
\textbf{Conditional Distributionally Robust Functional} \cite{shapiro2022}:  
Let \( (\Omega, \mathcal{F}) \) be a measurable space, and let \( \mathcal{M} \) be an ambiguity set of probability measures on \( (\Omega, \mathcal{F}) \). For a random variable \( Z: \Omega \to \mathbb{R} \), the conditional distributionally robust functional is defined as  
\begin{equation}\label{eq:condn_drf_defn}
    \mathcal{R} :=
    \sup_{\mathbb{P} \in \mathcal{M}} \mathbb{E}_{\mathbb{P}|\mathcal{F}}[Z],
\end{equation}  
where \( \mathbb{E}_{\mathbb{P}|\mathcal{F}}[Z] \) denotes the conditional expectation of \( Z \) given the \( \sigma \)-algebra \( \mathcal{F} \).  
\end{definition}

The conditional distributionally robust functional, as defined in Definition~\ref{def: condn_drf}, serves as the core component for constructing a distributionally robust risk measure. This functional allows us to quantify the risk of a random variable in the worst-case scenario, where the uncertainty in the underlying probability distribution is captured by the ambiguity set $\mathcal{M}$.

With this framework in place, we are now in a position to define and quantify our distributionally robust risk measure. 

\section{Distributionally Robust Risk of Cascading Failures}  \label{risk}
In this section, we present a framework for quantifying Distributionally Robust (DR) Cascading Risk, focusing on quantification of ambiguity in probability measures, the sigma algebra associated with observables, and the assessment of risk within this context.
\subsection{Quantification of Ambiguity Set}
The ambiguity in the probability distribution arises from three primary sources: time delay, diffusion coefficient, and network fluctuations as evident from Lemma \ref{lem:sigma_y_steady}. These combined sources of uncertainty result in a distribution that deviates from the nominal model. In this work, we focus specifically on the uncertainty introduced by the diffusion coefficient and its impact on the resulting probability distribution.


\begin{assumption}\label{asmp:ambiguity_set_input}  
Since the network dynamics \eqref{eqn: network_dynamics} is driven by Brownian motion, we define the ambiguity set of the input noise as the family of multivariate normal distributions with zero mean and an uncertain covariance matrix \(\Gamma = BB^T\):  
\begin{equation}\label{eqn:ambiguity_set_quantification_input}  
    \mathcal{M}_{\Gamma} = \left\{ \mathbb{P} \sim \mathcal{N}(0, \Gamma) \mid  (1- \varepsilon) \Gamma_0\preceq \Gamma \preceq \Gamma_0 (1 + \varepsilon) \right\},  
\end{equation}  
where \(\Gamma_0 = B_0 B_0^T\) is the nominal input noise covariance matrix corresponding to the estimated diffusion coefficient \(B_0\), and \(\varepsilon \in [0,1)\) (assumed known a priori) represents the radius of the ambiguity set, quantifying the level of uncertainty or deviation from the nominal covariance matrix.

\end{assumption}

\begin{lemma}\label{lem:ambiguity_B}
For the system given by \eqref{eqn: network_dynamics} with ambiguity set of the input noise given by assumption \ref{asmp:ambiguity_set_input},
the ambiguity set of the observables \(\bm{y}_t\) satisfies 
\begin{equation}\label{eqn:ambiguity_B_y}
    \mathcal{M}_{\Sigma} = \left\{ \mathbb{P} \sim \mathcal{N}(0, \Sigma) \mid  (1- \varepsilon) \Sigma_0\preceq \Sigma \preceq \Sigma_0 (1 + \varepsilon) \right\},  
\end{equation}
where \(\Sigma_0\) is the nominal covariance matrix of the observables.
\end{lemma}


\begin{assumption} \label{asmp:ambiguity_set_B}
    For the system \eqref{eqn: network_dynamics}, the diffusion coefficient satisfies \(B = bI_n\). As a result, the input noise covariance satisfies
    \begin{equation}\label{eqn:ambiguity_set_B}
        b_0^2(1-\varepsilon)\leq b^2 \leq b_0^2(1+\varepsilon),
    \end{equation}
\end{assumption}


From Lemma \ref{lem:sigma_y_steady} and Assumption \ref{asmp:ambiguity_set_B}, it follows that ambiguity set of observables can be simplified as 
        \begin{equation}\label{eqn:ambiguity_set_b}
   \bar{\mathcal{M}}_{\Sigma} = \left\{\mathbb{P} \sim \mathcal{N}(0, \Sigma) \mid \lvert \frac{\sigma_{ij} - \sigma_{ij,0}}{\sigma_{ij,0}}\rvert \leq \varepsilon\right\} ,
    \end{equation}
where \(\sigma_{ij,0} =\Sigma_0(i,j).\) and \(\sigma_{ij} =\Sigma(i,j).\)

\subsection{Quantification of DR Cascading Risk}
In this subsection, we focus on the quantification of distributionally robust (DR) risk in the context of cascading failures in multi-agent rendezvous. We aim to derive analytical expressions for DR risk, considering  the ambiguity in probability distributions due to inherent uncertainty in diffusion coefficient as stated in \eqref{eqn:ambiguity_set_b}.

\noindent Let us assume that the agents are labeled as \( \{1, \dots, n\} \). The main scenario considers the failure of the \( i \)th agent to achieve \( c \)-consensus, characterized by the condition \( |y_i| > c \). Given this failure event, we analyze the distributionally robust risk of failure for the \( j \)th agent.  

To quantify the risk of cascading failures, we define the failure event in achieving \( c \)-consensus for the observable \( \bar{y}_{i} \) as
\begin{align*}
    \Big\{ \bar{y}_{i} \in U_{\delta^i} \Big\} \text{ with } U_{\delta^i}=\left( -\infty, -\delta^i - c\right) \bigcup \left(  \delta^i+c  ,\infty \right),
\end{align*}
where $c, \delta^i \in \R_{+}$ and \(\delta^i\) quantifies the magnitude of deviation of observable \(y_i\). 
 \\
We denote the sigma-algebra generated by a particular superset ${U}_{\delta^i}$ as $\mathcal{F}^i = \sigma\left(\mathcal{U}_{\delta^i}\right)$ such that $\mathcal{F}^i = \{\phi, \mathcal{U}_{\delta_i}, \mathcal{U}_{\delta^i}^k, \R\}$, where $\mathcal{U}_{\delta^i}^k$ is the complement of the set $\mathcal{U}_{\delta^i}$. Further, we denote the \(\sigma-\) trivial algebra as $\mathcal{F}^{i,o} = \{\phi, \R\}.$

To quantify risk, we first express the formula for conditional expectation in a more familiar form, as shown below.
\begin{lemma}\label{lem:conditional_expectation}
    The conditional expectation of random variable \(\lvert y_j \rvert\) given \(y_i \in U_{\delta^i}\) can be written as
    \begin{equation}\label{eqn:conditional_expectation}
        \mathbb{E}_{\mathbb{P}|\mathcal{F}^i}[ \lvert y_j \rvert ] = \dfrac{\mathbb{E}_{\mathbb{P}}[ \lvert y_j \rvert \bm{1}_{y_i \in U_{\delta^i}}]}{\mathbb{P}[ {y_i \in U_{\delta^i}}]}
    \end{equation}
\end{lemma}

\noindent
This lemma serves as a critical step in quantifying the risk of cascading failures in multi-agent rendezvous, particularly by defining the conditional expectation of agent \( j \)’s deviation, given the information available from agent \( i \)’s observable. This formulation captures the inter dependencies between agents, enabling a deeper understanding of the cascading risk.

\noindent Building on the framework outlined in \cite{Peyman2018, shapiro2022}, we now introduce the distributionally robust risk measure for cascading failures. Specifically, we define the distributionally robust cascading risk measure as follows:

\begin{equation}\label{eqn:condn_dr_risk_definition}
    \mathcal{R}^j_i := 
    \underset{\mathbb{P} \in \bar{\mathcal{M}}_{\Sigma}}{{\sup}} \left( \underset{\delta}{\text{inf}}~\mathbb{E}_{\mathbb{P}|\mathcal{F}^i}[\lvert y_j \rvert] \in U_{\delta^{+}} \right) ,
\end{equation}
where \(U_{\delta^{+}} = \left(\delta+c  ,\infty \right).\) The measure \( \mathcal{R}^j_i \) in \eqref{eqn:condn_dr_risk_definition} quantifies the worst-case risk of failure for agent \(j\) given agent \(i\) has already failed by taking into account the ambiguity set defined in \eqref{eqn:ambiguity_set_b}.

For the exposition of our next result, we introduce following notations:
\[\mathbb{E}^j_i = \mathbb{E}_{\mathbb{P}|\mathcal{F}^i}\left[\lvert y_j \rvert ~ \right], \hspace{0.5cm} h(\cdot) = \textnormal{erf}(\cdot), \hspace{0.5cm}\bar{\delta} = \delta^i + c.\]



\begin{theorem}\label{thm:DR_risk}
Suppose the system \eqref{eqn: network_dynamics} has reached a steady state, and agent \(i\) has failed to achieve \(c\)-consensus, with its observable \(y_i\) belonging to the uncertainty set \(U_{\delta^i}\). The distributionally robust cascading risk of agent \(j\) is given by  
\begin{equation}\label{eqn:DR_risk}
    \mathcal{R}^j_i =
    \begin{cases}
       0, & \text{if} ~ \underset{\mathbb{P} \in \bar{\mathcal{M}}_{\Sigma}}{\sup}~ \mathbb{E}^j_i \leq c, \\
        \underset{\mathbb{P} \in \bar{\mathcal{M}}_{\Sigma}}{\sup}~ \mathbb{E}^j_i - c, & \text{otherwise},
    \end{cases}
\end{equation}
where \(\mathbb{E}^j_i\), the expected value of \(y_j\) given that \(y_i \in U_{\delta^i}\), is given by  
\begin{equation}\label{eqn:conditional_expectation_closed_form}  
        \mathbb{E}^j_i =  \sqrt{\frac{2}{\pi}} \frac{\sigj}{1-h(\delta^{*})} \left[1 - h\left(\frac{\delta^{*}}{\rho'}\right) +\rho h\left(\frac{\rho \delta^{*}}{\rho'}\right)\textnormal{e}^{-{\delta^{{*}^2}}}\right],
\end{equation}  
where \({\delta^{*}} = \frac{\bar{\delta}}{\sqrt{2}\sigi}\), and \(\rho, \rho'\) are as defined in Lemma \ref{lem:bivariate_normal}.  
\end{theorem}
The two cases of distributionally robust cascading risk are straightforward: \(\mathcal{R}^j_i = 0\) ensures that the expected observable of agent \(j\) remains within tolerance \(c\), while a nonzero risk implies failure to reach \(c\)-consensus in expectation. From \eqref{eqn:DR_risk}, risk depends on time delay, graph topology, and agent \(i\)'s failure mode, quantified by \(\bar{\delta}\). A higher expectation leads to greater risk, enabling network design to minimize failure probability.

Taking the supremum over all distributions in the ambiguity set ensures a worst-case risk assessment, providing a robust measure of vulnerability to cascading failures.



The risk vector of all agents can then be written as 
\begin{equation}\label{eqn:risk_vector}
    \bm{ \mathcal{R}}^j_i = \left[ \mathcal{R}^1_i, \dots,  \mathcal{R}^n_i,\right]
\end{equation}
where \(\mathcal{R}^j_i = 0\) for \(i = j.\)

An immediate consequence of Theorem \ref{thm:DR_risk} is the case of independent agents.  

\begin{corollary}\label{cor:single_risk_rho_0}
If agents \(j\) and \(i\) are uncorrelated, the distributionally robust cascading risk of agent \(j\)reduces to the single-agent failure risk:  
\begin{equation}\label{eqn:DR_risk_single}
        \mathcal{R}^j = \underset{\mathbb{P} \in \bar{\mathcal{M}}_{\Sigma}}{\textnormal{sup}}~ \mathbb{E}_{\mathbb{P}}\left[\lvert y_j\rvert\right] - c = \sqrt{\frac{2}{\pi}} \sigj(1+ \varepsilon) - c.
    \end{equation}
\end{corollary}  

This result quantifies the inherent failure risk of a single agent, corresponding to the case where agent \(i\)'s information is captured by the trivial \(\sigma\)-algebra \(\mathcal{F}^{i,o}\).

Next, we discuss how network affects risk quantification, formalizing this with an example of a complete graph where all edges have uniform weights.

\begin{lemma}\label{lem:dr_risk_complete}
For a complete graph with uniform edge weights \(\omega\), the distributionally robust risk of agent \(j\), given the observable of agent \(i\), \(y_i \in U_{\delta^i}\), is a function of the number of agents \(n\) and edge weights \(\omega\), expressed as:
\begin{equation}\label{eqn:DR_risk_complete}
\mathcal{R}^j_i =
\begin{cases}
0, & \text{if} ~ \underset{\mathbb{P} \in \bar{\mathcal{M}}_{\Sigma}}{\sup}~ \mathbb{E}^j_{i,f} \leq c, \\
\underset{\mathbb{P} \in \bar{\mathcal{M}}_{\Sigma}}{\sup}~ \mathbb{E}^j_{i,f} - c & \text{otherwise},
\end{cases}
\end{equation}
where
\begin{equation}\label{eqn:cond_expect_complete_graph}  
\mathbb{E}^j_{i,f} =  \sqrt{\frac{2}{\pi}} \frac{f(\omega n \tau)}{h^c({\delta^{*}_f})} \left[h^c\left({{\delta^{*,\dag}_f}}\right) +\frac{1}{n-1}h\left({{\delta}^{*,\ddag}_f}\right)\textnormal{e}^{-{{\delta}_f^{*^2}}}\right],
\end{equation}  
and \(h^c(\cdot) = 1 - h(\cdot)\), \(f(\omega n \tau) = \frac{b^2 \cos(\omega n \tau)}{2 \omega n \tau (1 - \sin(\omega n \tau))}\), \(\delta^{*}_f = \frac{\bar{\delta}}{\sqrt{2} f(\omega n \tau)}\), \(\delta_f^{*,\dag} = \frac{(n-1) \bar{\delta}}{f(\omega n \tau) \sqrt{n(n-2)}}\), and \(\delta_f^{*,\ddag} = \frac{\bar{\delta}}{f(\omega n \tau) \sqrt{n(n-2)}}\).
\end{lemma}

Lemma \ref{lem:dr_risk_complete}, together with Lemma \ref{lem:Sigma_L_monotonicity}, is used to analyze the impact of network connectivity on risk, which we explore in the next section.

\section{Case Studies}

We consider a team of \( n = 21 \) agents, where the parameters of the consensus model \eqref{eqn: network_dynamics} are set as \( \tau = 0.05 \), \( b_0 = 4 \), and \( c = 0.1 \). In all simulation studies, we assume that the observable of agent \( i = 11 \) lies outside the range \( |c| \), specifically within the set \( U_{\delta^i} \).

We investigate case studies of the rendezvous problem, focusing on consensus dynamics governed by equation (3) under three different communication topologies: complete, path, and \( p \)-cycle graphs \cite{van2010graph}.



\begin{figure}[t]
    \begin{subfigure}[t]{.48\linewidth}
        \centering
    \includegraphics[width=\linewidth]{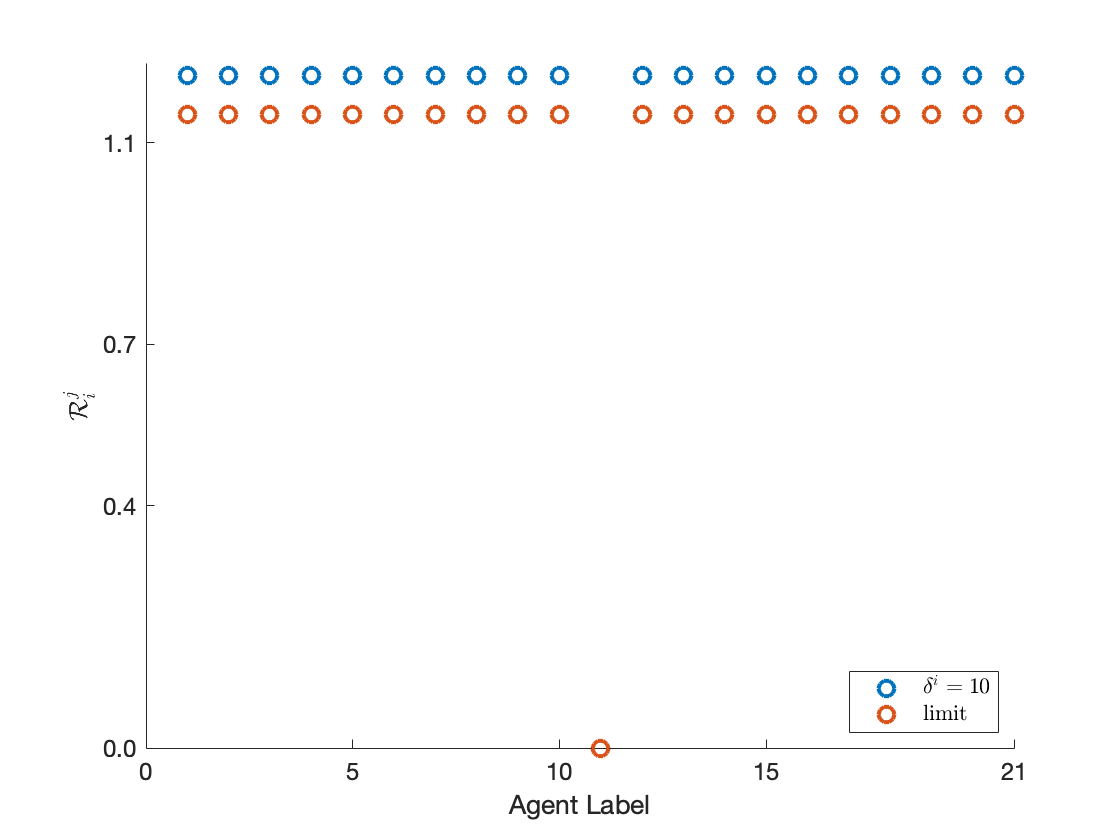}
    	\caption{The complete graph.}
    \end{subfigure}
    \hfill
    \begin{subfigure}[t]{.48\linewidth}
        \centering
    \includegraphics[width=\linewidth]{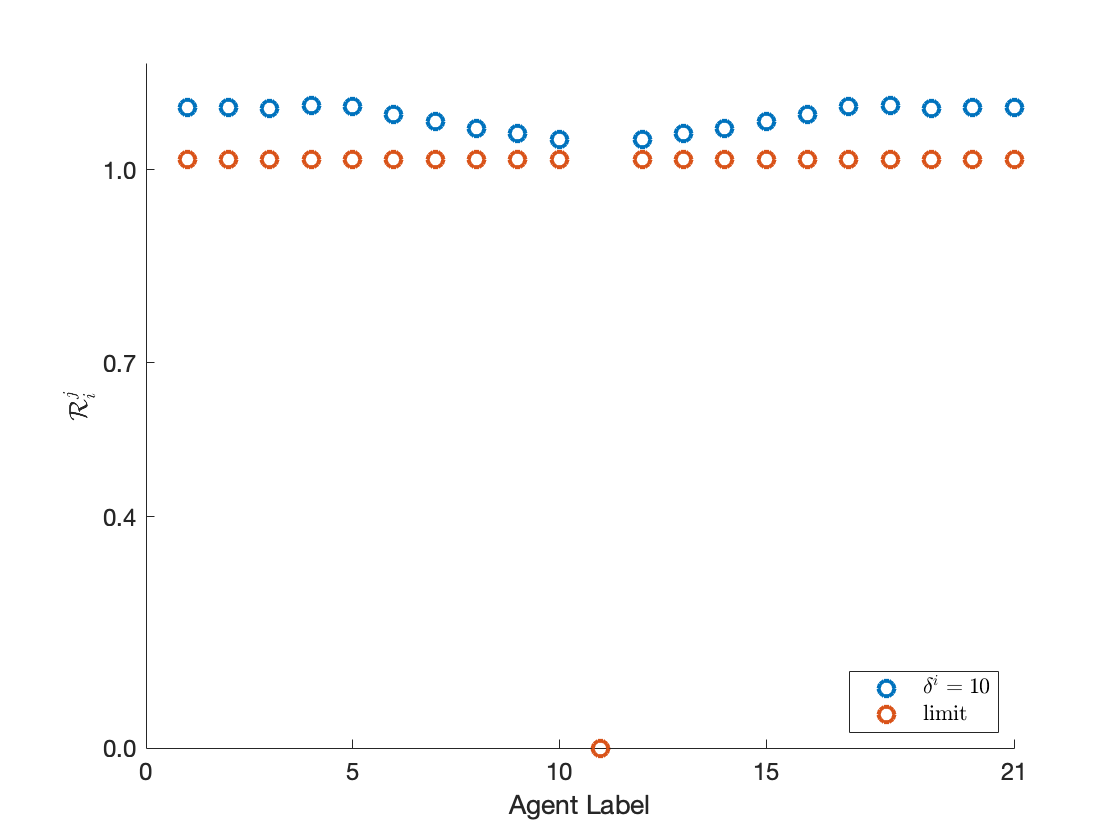}
    	\caption{The $14-$cycle graph.}
    \end{subfigure}
    \hfill
    \begin{subfigure}[t]{.48\linewidth}
        \centering
\includegraphics[width=\linewidth]{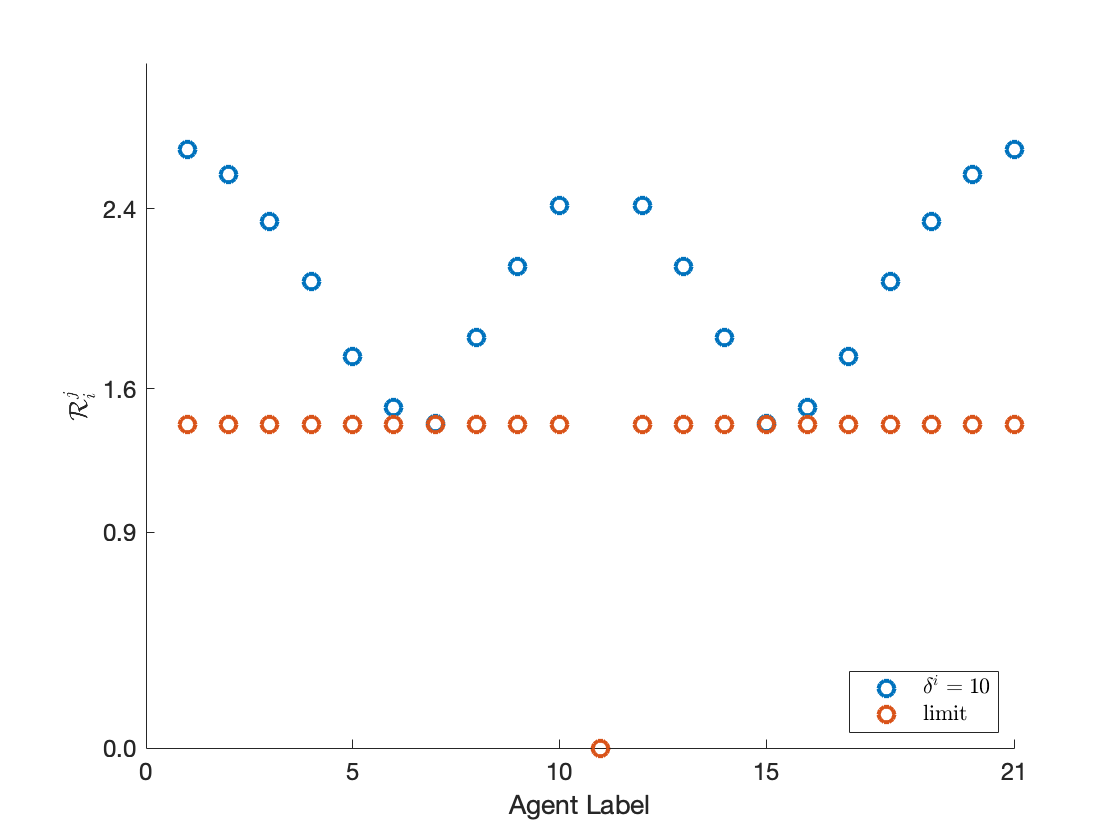}
    	\caption{The $6-$cycle graph.}
    \end{subfigure}
    \hfill
    \begin{subfigure}[t]{.48\linewidth}
        \centering
    \includegraphics[width=\linewidth]{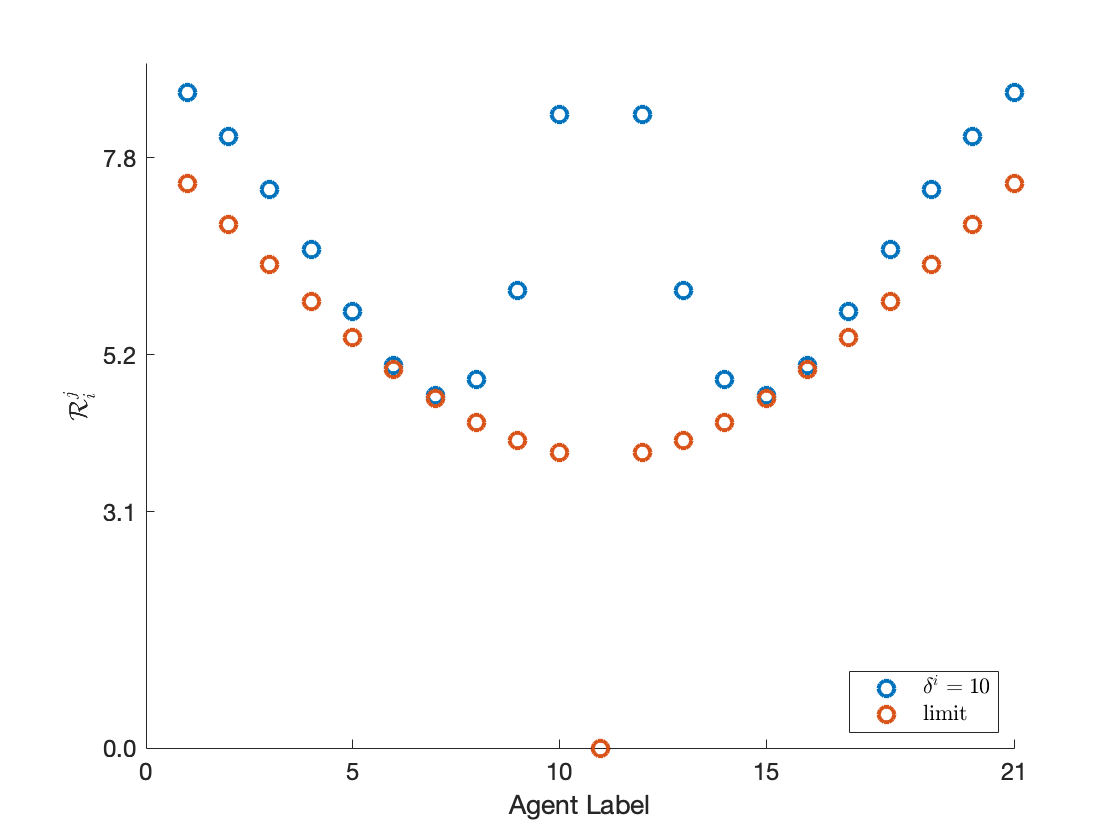}
    	\caption{The path graph.}
    \end{subfigure}
    \hfill
    \caption{The distributionally robust \(\left(\varepsilon = 0.3\right)\) cascading risk profile for various graph topologies. }
    \label{fig:risk_delta_10}
\end{figure}
\subsection{DR Risk of Cascading Failure}

The distributionally robust risk profile of cascading failure in achieving \( c \)-consensus is evaluated using the closed-form expression from Theorem \ref{thm:DR_risk} across different unweighted communication graph topologies (edge weights set to 1), as shown in Fig.~\ref{fig:risk_delta_10}. 
For the complete graph, the risk profile remains uniform across all agents, consistent with Lemma \ref{lem:dr_risk_complete}. In contrast, the results for path and \( p \)-cycle graphs illustrate the significant influence of network topology on the relative risk of individual agents.

\begin{figure*}[t]
    \centering
    \begin{subfigure}[t]{.24\linewidth}
        \centering
        \includegraphics[width=\linewidth]{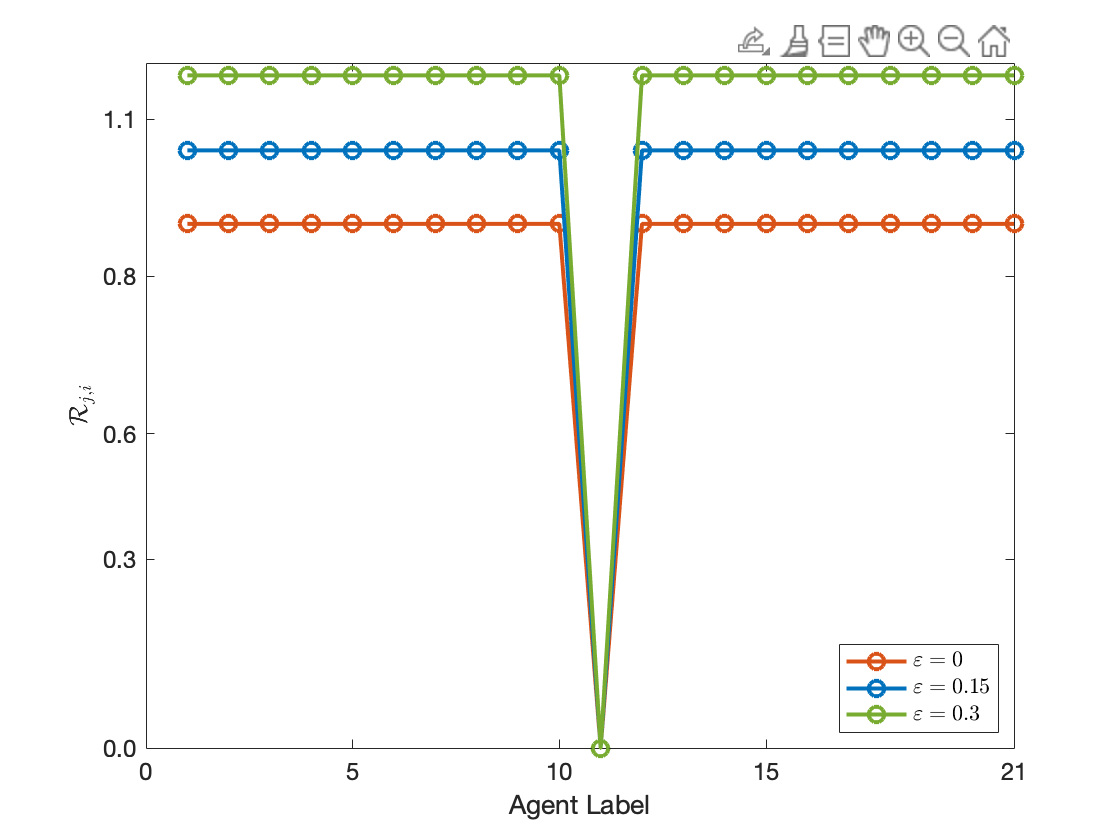}
        \caption{The complete graph.}
    \end{subfigure}
    \hfill
    \begin{subfigure}[t]{.24\linewidth}
        \centering
        \includegraphics[width=\linewidth]{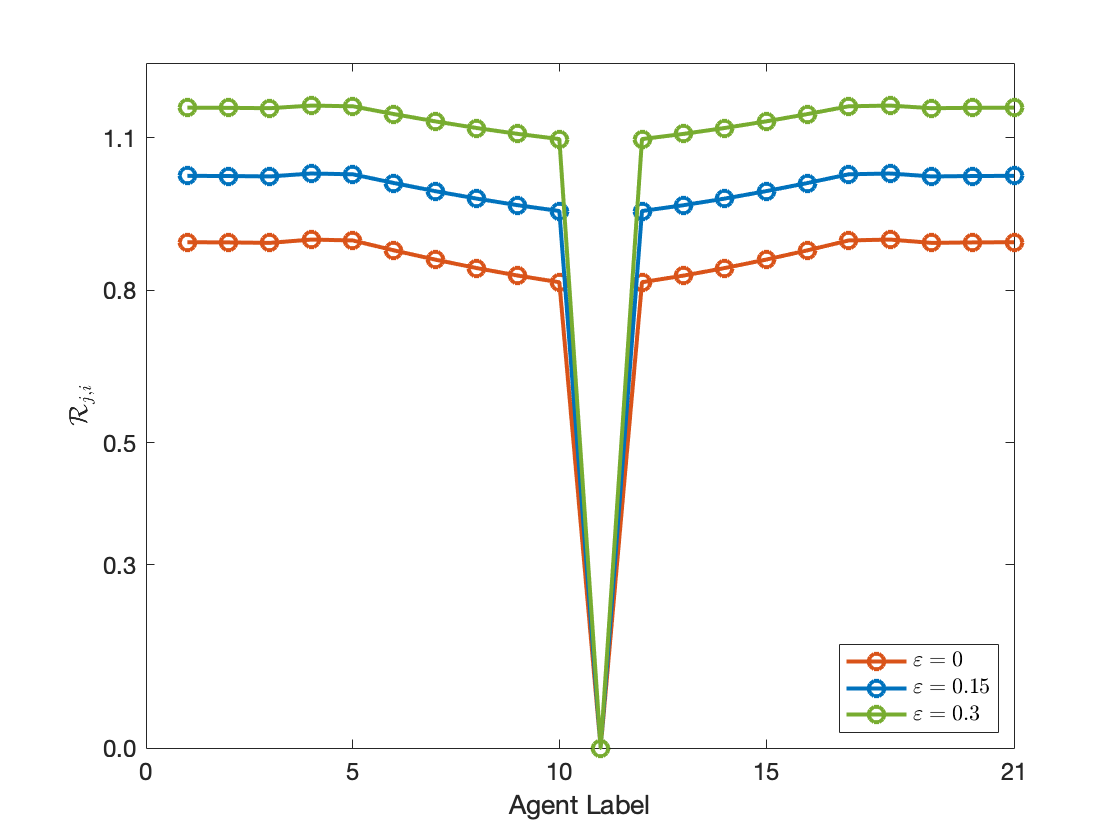}
        \caption{The $14$-cycle graph.}
    \end{subfigure}
    \hfill
    \begin{subfigure}[t]{.24\linewidth}
        \centering
        \includegraphics[width=\linewidth]{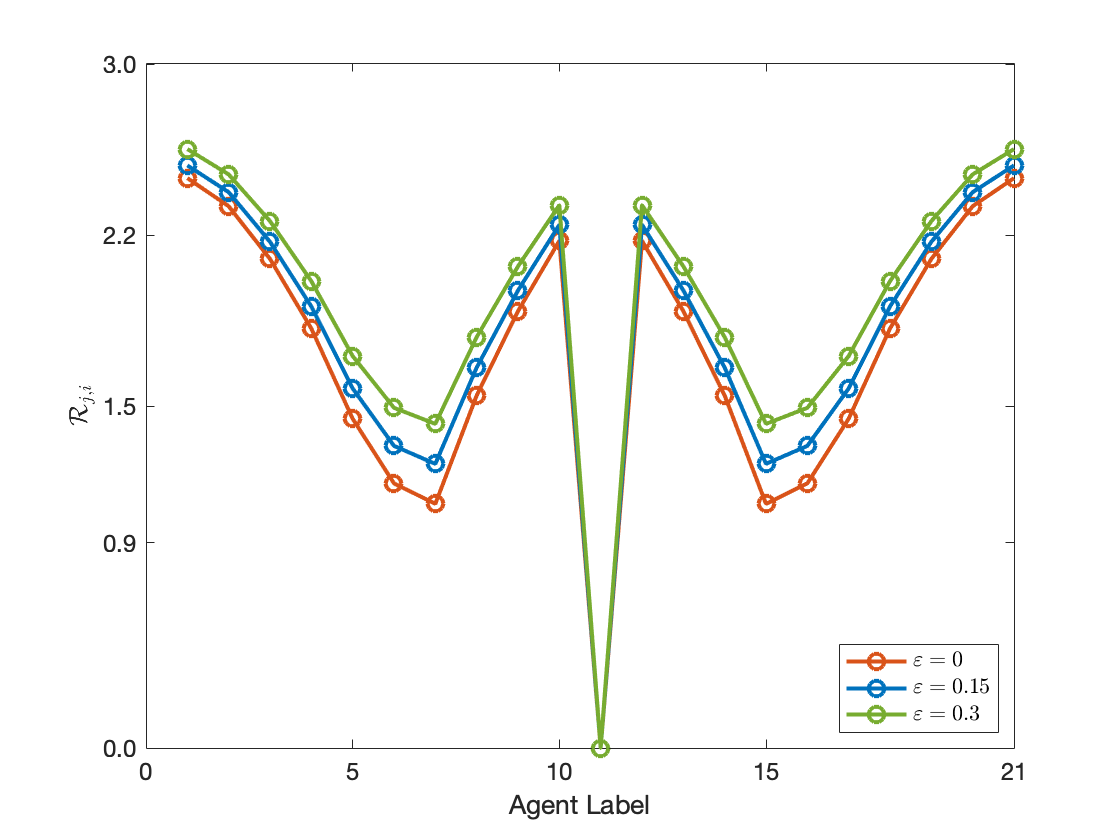}
        \caption{The $6$-cycle graph.}
    \end{subfigure}
    \hfill
    \begin{subfigure}[t]{.24\linewidth}
        \centering
        \includegraphics[width=\linewidth]{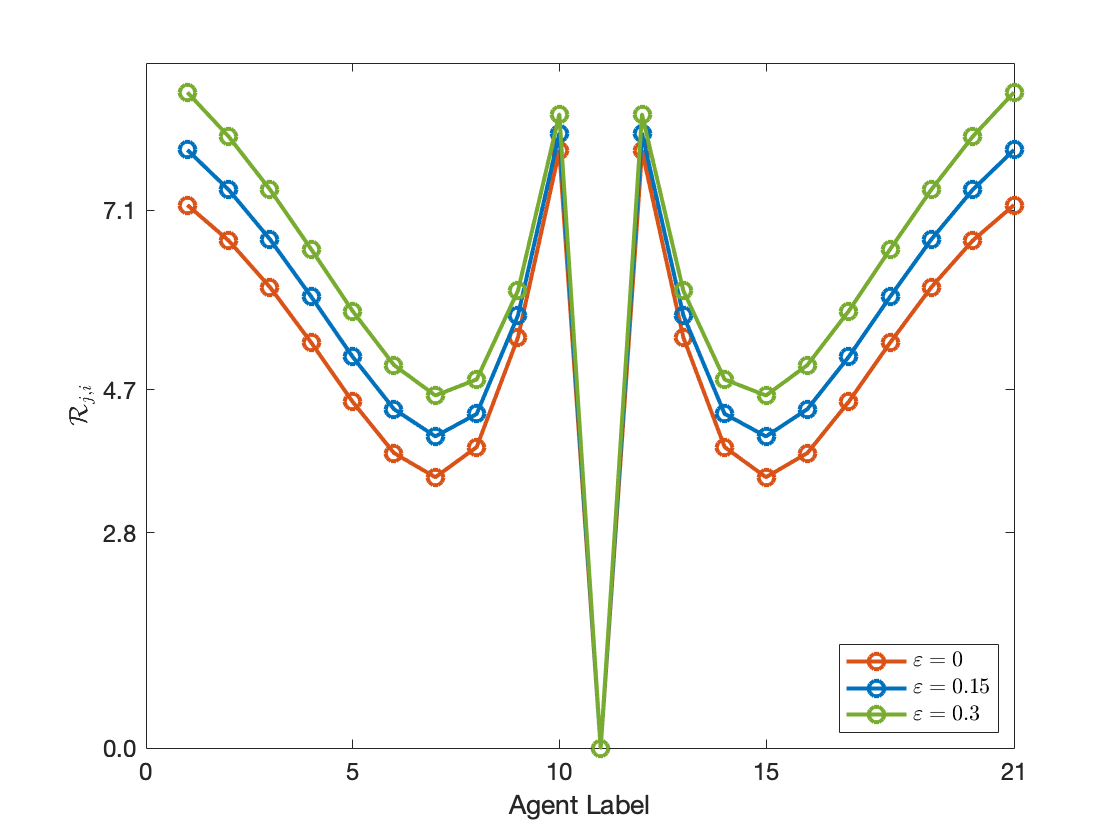}
        \caption{The path graph.}
    \end{subfigure}
    \caption{The distributionally robust cascading risk profile for different graph topologies for various value of \(\varepsilon\).}
    \label{fig:risk_epsilon}
\end{figure*}

Furthermore, as shown in Fig.~\ref{fig:risk_epsilon}, increasing the size of the ambiguity set leads to a higher distributionally robust cascading risk, which is an expected outcome.


\subsection{Impact of Network Connectivity on DR Risk}
\begin{figure}[h]
    \centering
    \begin{subfigure}[t]{0.48\linewidth}
        \centering
        \includegraphics[width=\linewidth]{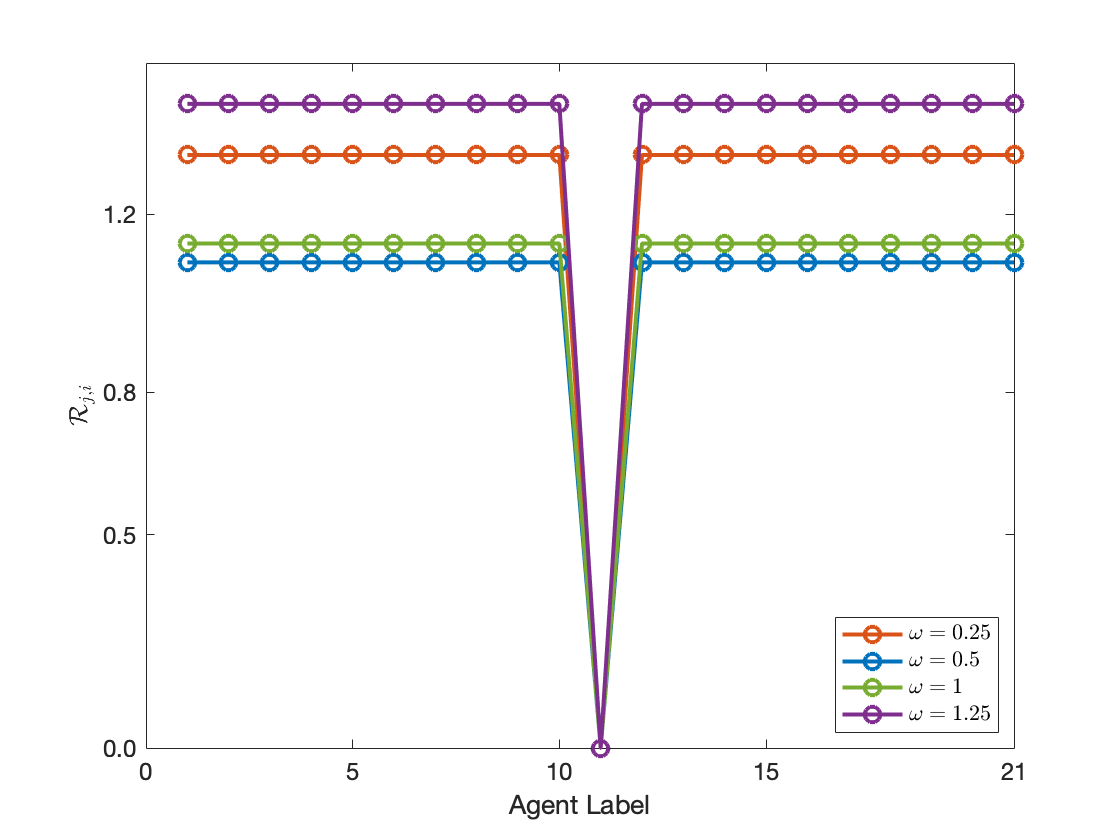}
        \caption{The complete graph.}
    \end{subfigure}
    \hfill
    \begin{subfigure}[t]{0.48\linewidth}
        \centering
        \includegraphics[width=\linewidth]{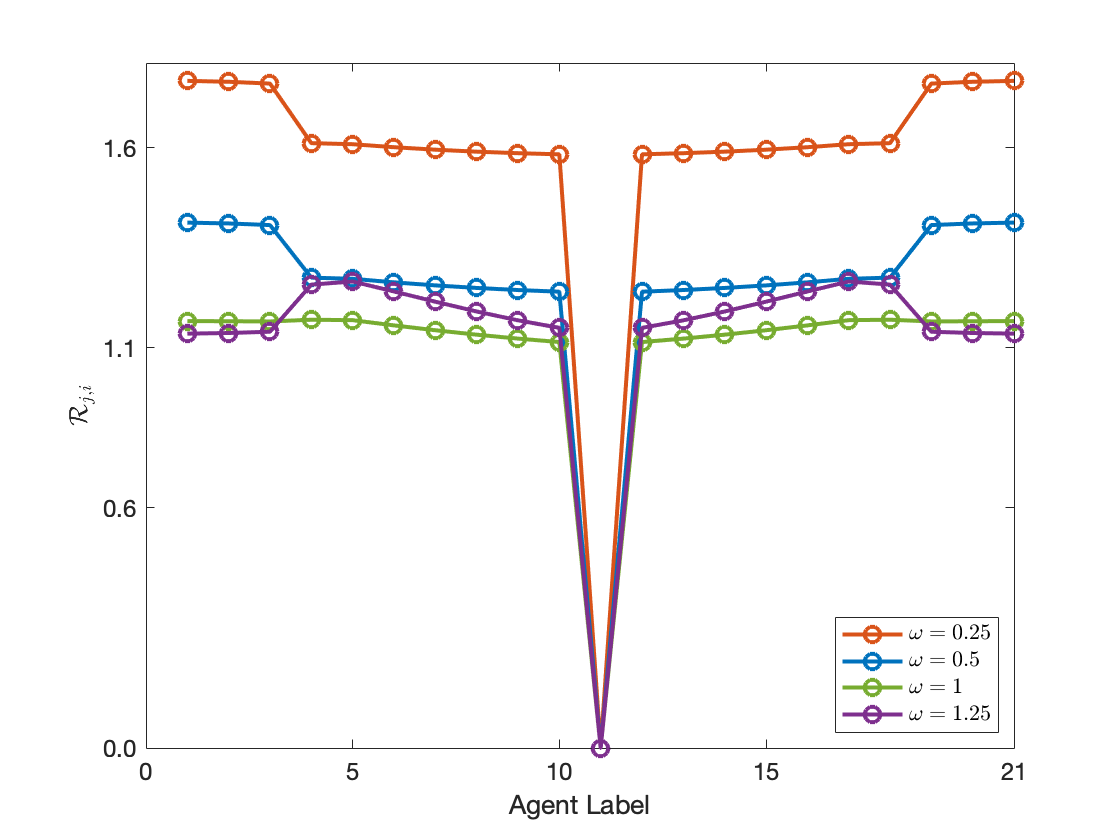}
        \caption{The $14$-cycle graph.}
    \end{subfigure}
    \vfill
    \begin{subfigure}[t]{0.48\linewidth}
        \centering
        \includegraphics[width=\linewidth]{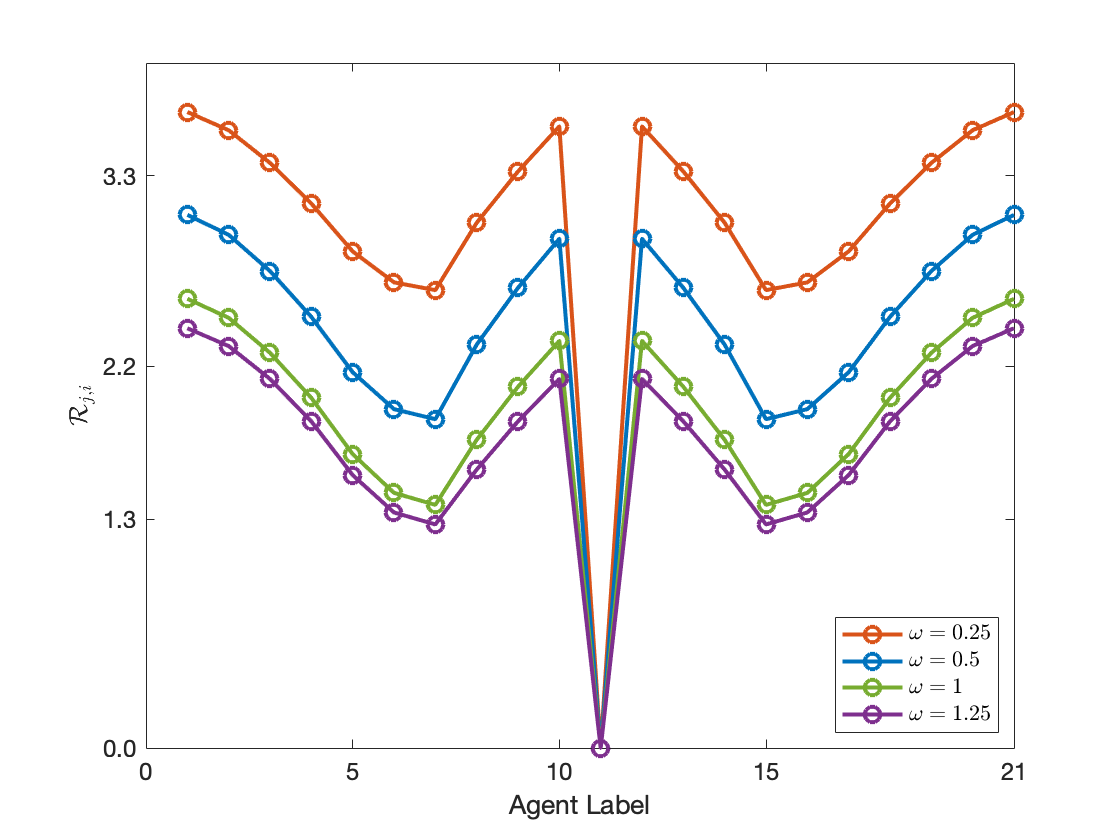}
        \caption{The $6$-cycle graph.}
    \end{subfigure}
    \hfill
    \begin{subfigure}[t]{0.48\linewidth}
        \centering
        \includegraphics[width=\linewidth]{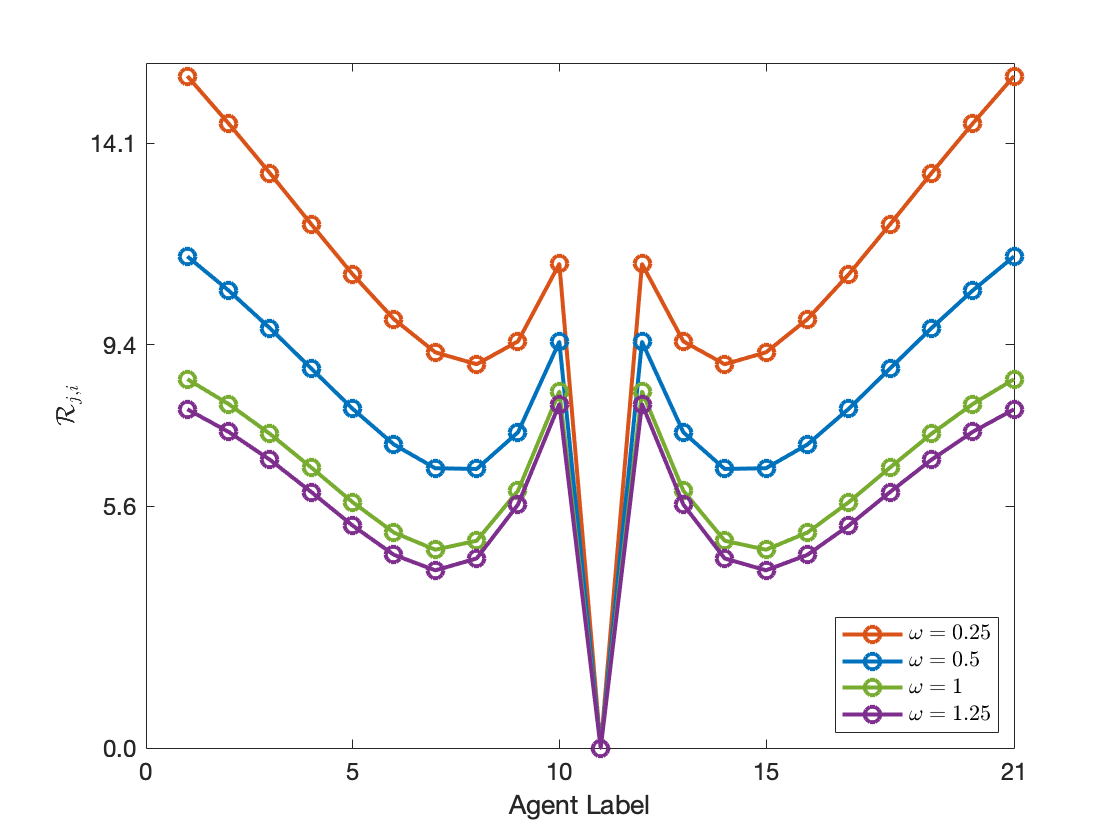}
        \caption{The path graph.}
    \end{subfigure}
    \caption{The distributionally robust (\(\varepsilon = 0.3\)) cascading risk profile  for various graph topologies with uniform change in edge weights.}
    \label{fig:risk_vs_weight}
\end{figure}

While increased network connectivity is generally associated with improved robustness, it does not always result in a reduction in distributionally robust cascading risk. This phenomenon is illustrated in Fig.~\ref{fig:risk_vs_weight}, particularly in the cases of the complete graph and the p-cycle graph, where the risk exhibits a certain degree of non-monotonicity with respect to the edge weights. This behavior can be understood through the dependence of the covariance matrix on the graph Laplacian, as stated in Lemma~\ref{lem:Sigma_L_monotonicity}, which quantifies the non-monotonic relationship between connectivity and the covariance matrix within the cone of positive semidefinite matrices. Additionally, the closed-form risk expression for the complete graph, given in Lemma~\ref{lem:dr_risk_complete}, further supports this observation. Our results highlight that increasing connectivity in certain regions (Fig.~\ref{fig:fx_graph}) can, counterintuitively, elevate the risk of cascading failures.


\section{Conclusion}
This work develops a distributionally robust risk framework for analyzing cascading failures in time-delayed, first-order networked dynamical systems. We derive explicit formulas for the ambiguity set and cascading risk, offering theoretical insights and empirical validation. Our results show how steady-state statistics shape risk in multi-agent rendezvous, highlighting the critical role of ambiguity sets and the nontrivial impact of network connectivity on risk.  

Future work will focus on conditioning risk on multiple failures, extending the framework to broader class of ambiguity sets, and exploring how network properties like connectivity and effective resistance constrain minimum risk. These insights will aid in designing resilient distributed control strategies.



\appendix
\subsubsection{Proof of Lemma \ref{lem:sigma_y_steady}}
A detailed proof of this result is provided in \cite{Somarakis2019g}.
\hfill$\square$

\subsubsection{Proof of Lemma \ref{lem:Sigma_L_monotonicity}}

 From graph of \(f(x) = \frac{\cos(x)}{2x(1 - \sin(x))}\) in Fig. \ref{fig:fx_graph} and Assumption \ref{asp:stable}, it is clear that \(\frac{\cos (\lambda_k \tau)}{\lambda_k (1 - \sin (\lambda_k \tau))}\) is positive for all \(k \in \{2, \dots, n\} \) and decreasing if \(\lambda_n \tau \leq \bar{\lambda} \tau\). For two graph Laplacians \(L_1\) and \(L_2\)  with same graph topology and uniform edge weights \(\omega_1\) and \(\omega_2\) such that \(\omega_1 \leq \omega_2\) with \(\lambda_{n,2} \tau \leq \bar{\lambda}\tau,\) it is easy to conclude \(\bar{\Lambda}^1 \succeq \bar{\Lambda}^2\) as \(\frac{\cos (\lambda_{k,1} \tau)}{\lambda_{k,1} (1 - \sin (\lambda_{k,1} \tau))} \geq \frac{\cos (\lambda_{k,2} \tau)}{\lambda_{k,2} (1 - \sin (\lambda_{k,2} \tau))}\) for all \(k \in \{2, \dots, n\}.\) The result follows from the fact that \(L_1\) and \(L_2\) are simultaneously diagonalizable. The other inequality follows similarly.
\hfill$\square$

\subsubsection{Proof of Lemma \ref{lem:bivariate_normal}}
The proof follows from bi-variate normal distribution in \cite{tong2012multivariate} and completing the square in the exponent term.
\hfill$\square$
\subsubsection{Proof of Lemma \ref{lem:ambiguity_B}}
Using the result in \cite{Somarakis2017a}, the steady state covariance matrix of \(\bm{y}_t\) is given by 
\begin{align*}
    \Sigma = \lim_{t \to\infty}\int_{0}^{t} M_n  \Phi_L(t-s) B B^T \Phi_L^T(t-s)  M_n ds,
\end{align*}
    where \(\Phi_L(\cdot)\) is the principle solution of deterministic part of\eqref{eqn: network_dynamics}.
    For the inequality \(\Gamma = BB^T\succeq (1-\varepsilon)B_0B_0^T = (1-\varepsilon)\Gamma_0\), it is straightforward to verify that the following holds on the cone of positive semi-definite matrix.
    \begin{equation} \label{eqn:ambiguity_set_cone_y}
        \Tilde{\Phi}_L(t-s) \Gamma \Tilde{\Phi}_L^T(t-s)  \succeq (1-\varepsilon)
    \Tilde{\Phi}_L(t-s) \Gamma_0 \Tilde{\Phi}_L^T(t-s), 
    \end{equation}
where \(\Tilde{\Phi}_L(t-s) = M_n {\Phi}_L(t-s).\)
Taking the integral and limit on both sides of \eqref{eqn:ambiguity_set_cone_y} establishes the lower bound. The right inequality follows similarly.
\hfill$\square$
\subsubsection{Proof of Lemma \ref{lem:conditional_expectation}}
Using the notation in \cite{durrett2019probability}
\begin{align*}
    \mathbb{E}_{\mathbb{P}|\mathcal{G}_i}[\lvert y_j \rvert] &= \frac{\mathbb{E}_{\mathbb{P}|\mathcal{G}_i}[ \lvert y_j \rvert; U_{\delta^i}]}{\mathbb{P}\left(y_i \in U_{\delta^i} \right))}\\
    &= \dfrac{\mathbb{E}_{\mathbb{P}}[\lvert y_j \rvert\bm{1}_{y_i \in U_{\delta^i}}]}{\mathbb{P}[ {y_i \in U_{\delta^i}}]}.
\end{align*}

\subsubsection{Proof of Theorem \ref{thm:DR_risk}} 
To prove this theorem, we state the following integration result as Lemma \ref{lem:korotkov_integral}.
\begin{lemma}\label{lem:korotkov_integral}
The following integral identities, adapted from \cite{korotkov2020_error}, will be used in the subsequent analysis:
\begin{multline} \label{eqn:korotkov_integral_two}
\int_{0}^{\infty} z \, \textnormal{erf}\left(a_1 z + b_1\right) \exp\left(-a_2 z^2\right) \, dz 
= \frac{\textnormal{erf}\left(b_1\right)}{2 a_2} \\
+ \frac{a_1}{2 a_2 \sqrt{A}} \exp\left(- \frac{a_2 b_1^2}{A}\right) \left(1 - \textnormal{erf}\left(\frac{a_1 b_1}{\sqrt{A}}\right)\right),
\end{multline}

\begin{equation} \label{eqn:korotkov_integral_one}
\int_{0}^{\infty} z \, \exp\left(-a_2^2 z^2\right) \, dz = \frac{1}{2 a_2^2}.
\end{equation}
where \(a_1 \geq 0\), \(a_2 > 0\), and \(A = a_1^2 + a_2\).
\end{lemma}
Using Lemma \ref{lem:conditional_expectation}
\[
    \mathbb{E}\left[ \lvert y_j \rvert| y_i \in U_{\delta^i}\right] =  \frac{\mathbb{E}\left[ \vert y_j \rvert \bm{1}_{\left[y_i \in U_{\delta^i}\right]}\right]}{\mathbb{P}\left[\bm{1}_{\left[y_i \in U_{\delta^i}\right]}\right]}
\]


We analyze the numerator and denominator separately.
\begin{align} \label{eqn: conditional_expectation_denominator}
\mathbb{P}\left[y_i \in U_{\delta^i}\right] \notag
    &= \int_{U_{\delta^i}}^{} ~ \frac{1}{\sigi \sqrt{2 \pi}} \text{exp}\left(- \frac{y_i^2}{\sigi^2}\right) dy_i \notag\\
    & = 1 - \text{erf} \left({\delta^{*}}\right),
\end{align}
where \(\delta^{*}\) is as defined in Theorem \ref{thm:DR_risk}.

Using joint distributions of \(y_j\) and \(y_i\) in Lemma \ref{lem:bivariate_normal}, 
\begin{align*}
    \mathbb{E}\left[ \lvert y_j \rvert \bm{1}_{\left[y_i \in U_{\delta^i}\right]}\right] &= \int_{-\infty}^{\infty} \int_{y_i \in U_{\delta^i}}^{} ~ \lvert y_j \rvert p(y_j,y_i)  dy_j dy_i\\
         &= \frac{2}{\sqrt{2 \pi} \sigj} \int_{0}^{\infty} y_j ~\text{exp}\left(-\frac{y_j^2}{2\sigj^2} \right) I_i dy_j,
\end{align*}
where 
\begin{align*}
    I_i &= \frac{1}{\sqrt{2 \pi} \rho' \sigi} \int_{y_i \in U_{\delta^i}}^{}  \text{exp}\left(- \frac{1}{2 \rho'^2 \sigi^2}  \left({y_i} - \rho \frac{y_j \sigi}{\sigj}\right)^2\right) dy_i,\\
    &= 1 - \frac{1}{2}\left(\text{erf}\left(\frac{\bar{\delta} + \rho \frac{\sigi}{\sigj}y_j}{\sqrt{2}\rho'\sigi}\right) + \text{erf}\left(\frac{\bar{\delta} - \rho \frac{\sigi}{\sigj}y_j}{\sqrt{2}\rho'\sigi}\right)\right),
\end{align*}

For simplicity of notation, let us consider  \(\kappa_{\bar{\delta}}^{\pm} = \frac{\bar{\delta} \pm \rho \frac{\sigi}{\sigj}y_j}{\sqrt{2}\rho'\sigi}\).

\begin{equation}\label{eqn:conditional_expectation_numerator}
        \mathbb{E}\left[\lvert y_j \rvert\bm{1}_{\left[y_i \in U_{\delta^i}\right]}\right] = \frac{1}{\sqrt{2 \pi} \sigj}\left(I_1 -  I_2\right), 
\end{equation}
where 
\begin{align*}
    I_1&= 2\int_{0}^{\infty} y_j  ~\text{exp}\left(-\frac{y_j^2}{2\sigj^2} \right) dy_j =   \mathbb{E}_{\mathbb{P}}[\lvert y_j \rvert],\\
    I_2 &=\int_{0}^{\infty} y_j \left(\text{erf}\left(\kappa_{\bar{\delta}}^{+}\right) + \text{erf}\left(\kappa_{\bar{\delta}}^{-}\right)\right)~\text{exp}\left(-\frac{y_j^2}{2\sigj^2} \right) dy_j
 \end{align*}

Using the result from Lemma \ref{lem:korotkov_integral} and substituting \(a_2 = \frac{1}{2 \sigj^2}\) in \eqref{eqn:korotkov_integral_one}, \(I_1\) is given by 
\begin{equation}\label{eqn:I1_evaluation}
    I_1 =   2\sigj^2 \hspace{0.5cm} \mathbb{E}_{\mathbb{P}}\left[\lvert y_j\rvert\right] = \sqrt{\frac{2}{\pi}} \sigma_j.
\end{equation}

Consider the following notation
\[I_2^{\pm} =\int_{0}^{\infty}\lvert y_j \rvert \left(\text{erf}\left(\kappa_{{\bar{\delta}}}^{\pm}\right) \right)~\text{exp}\left(-\frac{y_j^2}{2\sigj^2} \right) dy_j.\]

 Substituting the following in \eqref{eqn:korotkov_integral_one} to solve for \(I_2^{\pm}\):
  \[z = y_j \hspace{0.4cm} a_1 = \frac{\rho}{\sqrt{2} \rho' \sigj}, \hspace{0.4cm} b_1^{\pm} = \pm\frac{\bar{\delta}}{\sqrt{2}\rho' \sigi}, \hspace{0.4cm} a_2 = \frac{1}{2\sigj^2},\]

\begin{equation}\label{eqn:I2_evaluation}
    I_2 = \sqrt{\frac{2}{\pi}} \sigj \left[\textnormal{erf}\left(\frac{\delta^{*}}{\rho'}\right) -\rho \textnormal{erf}\left(\frac{\rho \delta^*}{\rho'}\right)\textnormal{exp}\left(-{\delta^{*^2}}\right)\right]
\end{equation}
The conditions in Lemma \ref{lem:korotkov_integral} is trivially satisfied for \(a_2\). For \(a_1\), without loss of generality \(\rho\) can be assumed positive for solving the integral in Lemma \ref{lem:korotkov_integral}, although in our problem \(\rho \in \left[-1,1\right].\) This is due to the fact that 
\[\text{erf}\left(\kappa_{\bar{\delta}}^{+}\left(-\rho\right)\right) = \text{erf}\left(\kappa_{\bar{\delta}}^{-}\left(\rho\right)\right) \]
The result follows from \eqref{eqn: conditional_expectation_denominator}, \eqref{eqn:conditional_expectation_numerator},\eqref{eqn:I1_evaluation}, \eqref{eqn:I2_evaluation} and \eqref{eqn:condn_dr_risk_definition}.
\hfill$\square$

\subsubsection{Proof of Corollary \ref{cor:single_risk_rho_0}}  
From \eqref{eqn:I1_evaluation} in the proof of Theorem \ref{thm:DR_risk}, we obtain  
\[
\mathbb{E}_{\mathbb{P}}\left[\lvert y_j \rvert\right] = \sqrt{\frac{2}{\pi}} \sigma_j.
\]  
Substituting \(\rho = 0\) in Theorem \ref{thm:DR_risk} yields \(\rho' = 1\). The result follows directly by canceling the corresponding terms in the numerator and denominator and then taking the supremum over the ambiguity set defined in \eqref{eqn:ambiguity_set_b}.  
\hfill$\square$


\subsubsection{Proof of Lemma \ref{lem:dr_risk_complete}}  
To prove this lemma, we observe that the nontrivial eigenvalues of a complete graph with \(n\) nodes and uniform edge weights \(\omega\) are given by \(\omega n\). Applying Lemma \ref{lem:sigma_y_steady}, we immediately obtain \(\rho = \frac{-1}{n-1}\). The result follows by first computing \(\sigma_j\) and \(\sigma_i\) using Lemma \ref{lem:sigma_y_steady} and subsequently applying Theorem \ref{thm:DR_risk}.  
\hfill$\square$




\printbibliography

@book{van2010graph,
    title = {{Graph spectra for complex networks}},
    year = {2010},
    author = {Van Mieghem, Piet},
    publisher = {Cambridge University Press}
}

@book{durrett2019probability,
    title = {{Probability: theory and examples}},
    year = {2019},
    author = {Durrett, Rick},
    volume = {49},
    publisher = {Cambridge university press}
}

@book{tong2012multivariate,
    title = {{The multivariate normal distribution}},
    year = {2012},
    author = {Tong, Yung Liang},
    publisher = {Springer Science {\&} Business Media}
}

@article{dorfler2012synchronization,
  title={Synchronization and transient stability in power networks and nonuniform Kuramoto oscillators},
  author={Dorfler, F. and Bullo, F.},
  journal={SIAM Journal on Control and Optimization},
  volume={50},
  number={3},
  pages={1616--1642},
  year={2012},
  publisher={SIAM}
}

@article{acemoglu2015systemic,
  title={Systemic risk and stability in financial networks},
  author={Acemoglu, D. and Ozdaglar, A. and Tahbaz-Salehi, A.},
  journal={American Economic Review},
  volume={105},
  number={2},
  pages={564--608},
  year={2015}
}

@article{bertsimas1998air,
  title={The air traffic flow management problem with enroute capacities},
  author={Bertsimas, D. and Patterson, S. S.},
  journal={Operations research},
  volume={46},
  number={3},
  pages={406--422},
  year={1998},
  publisher={INFORMS}
}

@article{Peyman2018,
    title = {{Data- driven distributionally robust optimization using the Wasserstein metric: Performance guarantees and tractable reformulations }},
    year = {2018},
    journal = {Mathematical Programming},
    author = {Esfahani, P. Mohajerin and Kuhn, D.},
    pages = {115--166},
    volume = {171},
    publisher = {Springer New York LLC},
}

@article{shapiro2022,
    title = {{Conditional distributionally robust functional }},
    year = {2022},
    journal = {Available on Optimization Online},
    author = {Shapiro, A. and Pichler, A.},
}

@book{korotkov2020_error,
    title = {{Integrals Related to Error Function}},
    year = {2020},
    author = {Korotkov Nikolai , E. and  Korotkov Alexander , N},
    publisher = {Chapman and Hall}
}

@article{zhang2019robustness,
  title={Robustness of interdependent cyber-physical systems against cascading failures},
  author={Zhang, Y. and Ya{\u{g}}an, O.},
  journal={IEEE Transactions on Automatic Control},
  volume={65},
  number={2},
  pages={711--726},
  year={2019},
  publisher={IEEE}
}

@article{zhang2018cascading,
  title={Cascading failures in interdependent systems under a flow redistribution model},
  author={Zhang, Y. and Arenas, A. and Ya{\u{g}}an, O.},
  journal={Physical Review E},
  volume={97},
  number={2},
  pages={022307},
  year={2018},
  publisher={APS}
}

@inproceedings{somarakis2018risk,
  title={Risk of collision in a vehicle platoon in presence of communication time delay and exogenous stochastic disturbance},
  author={Somarakis, C. and Ghaedsharaf, Y. and Motee, N.},
  booktitle={2018 IEEE Conference on Decision and Control (CDC)},
  pages={4487--4492},
  year={2018},
  organization={IEEE}
}

@inproceedings{Somarakis2017a,
    title = {{Aggregate fluctuations in time-delay linear consensus networks: A systemic risk perspective}},
    year = {2017},
    booktitle = {Proceedings of the American Control Conference},
    author = {Somarakis, C. and Ghaedsharaf, Y. and Motee, N.}
}

@article{rockafellar2002conditional,
    title = {{Conditional value-at-risk for general loss distributions}},
    year = {2002},
    journal = {Journal of Banking and Finance},
    author = {Rockafellar, R. Tyrrell and Uryasev, Stanislav},
    number = {7},
    pages = {1443--1471},
    volume = {26}
}

@inproceedings{Somarakis2016g,
    title = {{Interplays Between Systemic Risk and Network Topology in Consensus Networks}},
    year = {2016},
    booktitle = {IFAC-PapersOnLine},
    author = {Somarakis, C. and Siami, M. and Motee, N.},
    number = {22},
    volume = {49}
}

@article{rockafellar2000optimization,
    title = {{Optimization of Conditional Value-at-Risk}},
    year = {1999},
    journal = {Portfolio The Magazine Of The Fine Arts},
    author = {Rockafellar, R. T. and Uryasev, S.},
    pages = {1--26},
    volume = {2}
}

@article{Somarakis2020power,
    title = {{Risk of Phase Incoherence in Wide Area Control of Synchronous Power Networks with Time-Delayed and Corrupted Measurements}},
    year = {2023},
    journal = {IEEE Transactions on Automatic Control},
    author = {Somarakis, C. and Liu, G. and Motee, N.},
    number = {12},
    volume = {68}
}

@inproceedings{pandey2023dr_risk_second_order,
    title = {{Quantification of Distributionally Robust Risk of Cascade of Failures in Platoon of Vehicles}},
    author = {Pandey, V. and Liu, G. and Amini, A. and Motee, N.},
    booktitle = {62nd IEEE Conference on Decision and Control (CDC)},
    pages={7401--7406},
    year={2023},
    organization={IEEE}
}

@inproceedings{liu2023cr_risk_first_order,
    title = {{Cascading Waves of Fluctuation in Time-delay Multi-agent Rendezvous}},
    author = {Liu, G. and Pandey, V.  and Motee, N.},
    booktitle = {2023 American Control Conference (ACC)},
    pages={4110--4115},
    year={2023},
    organization={IEEE}
}

@article{Somarakis2019g,
    title = {{Time-delay origins of fundamental tradeoffs between risk of large fluctuations and network connectivity}},
    year = {2019},
    journal = {IEEE Transactions on Automatic Control},
    author = {Somarakis, C. and Ghaedsharaf, Y. and Motee, N.},
    number = {9},
    volume = {64}
}

@inproceedings{liu2021risk,
    title={Risk of Cascading Failures in Time-Delayed Vehicle Platooning}, 
    author={G. Liu and C. Somarakis and N. Motee},
    year={2021},
    booktitle = {IEEE Conference on Decision and Control},
    % eprint={2109.01963},
    % archivePrefix={arXiv},
}

@article{ren2007information,
  title={Information consensus in multivehicle cooperative control},
  author={Ren, W. and Beard, R. W. and Atkins, E. M.},
  journal={IEEE Control systems magazine},
  volume={27},
  number={2},
  pages={71--82},
  year={2007},
  publisher={IEEE}
}

@article{olfati2007consensus,
  title={Consensus and cooperation in networked multi-agent systems},
  author={Olfati-Saber, R. and Fax, J. A. and Murray, R. M.},
  journal={Proceedings of the IEEE},
  volume={95},
  number={1},
  pages={215--233},
  year={2007},
  publisher={IEEE}
}

@article{olfati2004consensus,
  title={Consensus problems in networks of agents with switching topology and time-delays},
  author={Olfati-Saber, R. and Murray, R. M.},
  journal={IEEE Transactions on automatic control},
  volume={49},
  number={9},
  pages={1520--1533},
  year={2004},
  publisher={IEEE}
}
\end{document}